# Biological Systems as Heterogeneous Information Networks: A Mini-review and Perspectives

Koki Tsuyuzaki
Bioinformatics Research Unit, Advanced Center for Computing and Communication, RIKEN, Japan
koki.tsuyuzaki@riken.jp

Itoshi Nikaido
Bioinformatics Research Unit, Advanced Center for Computing and Communication, RIKEN, Japan
Single-cell Omics Research Unit, Center for Developmental Biology, RIKEN, Japan
itoshi.nikaido@riken.jp

## ABSTRACT

In the real world, most objects and data have multiple types of attributes and inter-connections. Such data structures are named "*Heterogeneous Information Networks*" (HIN) and have been widely researched.

Biological systems are also considered to be highly complicated HIN. In this work, we review various applications of HIN methods to biological and chemical data, discuss some advanced topics, and describe some future research directions.

## 1 INTRODUCTION

The smallest unit of life is a cell. Inside the cell membrane, based on the concept of central dogma [1], information from DNA is transcripted to RNA and successively translated to proteins (Fig. 1). In recent years, non-coding RNA and chemical modifications of histone proteins and DNA sequences have also been identified as an exception to the traditional concept [2]. Beyond cell exterior, many signals such as pathogen infection, experimental perturbation, drug ingestion, and lifestyle interfere with biological systems. Such changes in the ambient cellular environment affect the biological systems both quantitatively (e.g., change in the amount of RNA) and qualitatively (e.g., change in the base sequence of DNA), finally causing a change in phenotypes such as disease, morphology and physiological properties.

Although there is no comprehensive experimental method for measuring all systems at once, "*omics*" research [3] has been tackling the exhaustive measurement of systems by limiting a single type of biomolecule. For example, genome (gene + ome) studies or genomics, focus on only DNA sequences, and transcriptome (transcript + ome) studies, or transcriptomics, focus only on RNA sequences, and so on (Fig. 1). Such studies are becoming possible because of the rapid development of experimental equipment such as microarray, massively parallel DNA sequencing technology, and mass-spectrometry [3]. Omics studies are often performed to determine the association of two or three data domains and can be expressed in ER (Entity-Relationship) diagram (Fig. 2, left). For example, the study of DEG (Differentially Expressed Genes) [4] focuses on the association between gene expression-phenotypes; such research investigates what genes and phenotypes are linked at an instance-level (Fig. 2, right). Likewise, GWAS (Genome-wide Association Study) [5] focuses on the association between SNP (Single Nucleotide Polymorphism)-phenotype and so on [3-33]. The abbreviations of other omics studies are summarized in Appendix.

Although many analytical methods have been proposed for each biological data type, such methods have been highly domain-specific, and cross-domain studies (e.g., multi-omics) have still not been straightforward. Recently, "*Heterogeneous Information Networks*" (HIN) [34,35] data structures and the analytical methods have been applied to some biological problems and can form a unified framework for handling such complicated and inter-connected data. In this work, we review the various applications of HIN-based methods to biological and chemical data, discuss some advanced topics, and describe some future research directions.

The remaining sections are organized as follows. Section 2 describes the definition of graph that is used in this paper. Section 3 presents the HIN-based methodology used in previous papers. Section 4 describes the effectiveness of HIN for biological data. Section 5 introduces the concept of "*Guilt-by-Association*", which is important for analyzing biological data. Section 6 introduces previous applications of HIN-based methods. Section 7 describes some future research directions. Finally, Section 8 concludes this paper.

## 2 DEFINITIONS

In this section, we define the data structure described in this work.

### DEFINITION 1: Homogeneous Information Network

Consider a single entity in the ER diagram in Fig. 2 such as gene regulation network (GRN) [16]. Such a system can be defined as an undirected graph $G = (V, E)$, where V represents the set of nodes, and E is the set of edges on the graph. Because of the types of nodes $|V| = 1$ and the types of edges $|V| = 1$, such data structure is defined as a homogeneous information network. The graph can also be defined as a symmetric adjacency matrix (Fig. 2, right), where the number of rows and columns are both the number of instances of the gene expression and each element has 1 (linked) or 0 (not linked).

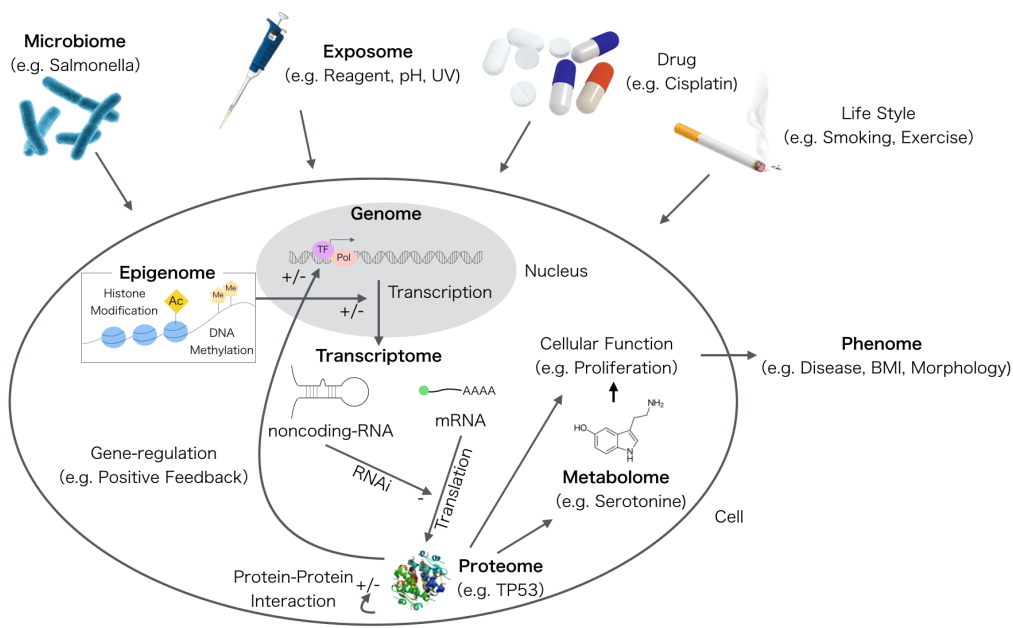

**Figure 1: Biological Systems and corresponding omics**

## DEFINITION 2: Heterogeneous Information Network

When the types of nodes or types of edges are larger than 1, the graph is defined as a "*Heterogeneous Information Network*" (HIN). For example, consider multiple entities connected to each other on the ER diagram in Fig. 2, such as DEG [4]. Although such systems are also defined as G' = (V', E'), where V' represents the set of nodes, E' represents the set of edges of the graph, |V'| = 2 and this graph is HIN. Such a relationship can also be defined as an asymmetric adjacency matrix in a similar manner of a homogeneous network (Fig. 2, right).

In this work, we introduce some HIN-based methods, while considering the difference of nodes. Although there are some methods for considering the difference of edge types, such data structures can form a higher-order matrix (*tensor*), which is a more advanced topic [36]. Therefore, we will focus on only the differences among node types.

## 3  METHODS USED FOR BIOLOGICAL HIN

In this section, we introduce the methods used on biological data defined as HIN. Although a comprehensive review of the algorithms is beyond the scope of this work, we will introduce some methods that are widely used on biological and chemical data. Most of the time, HIN-based methods are used for link prediction problems for biological HIN. After constructing the HIN by experimental data or external databases, many similarities between two nodes have been applied to decide which nodes are similar to each other on HIN.

Many similarity scores of graph theory and complex networks are applied to HIN data to capture local or global network topology, such as direct neighborhood, shortest path length, diffusion Kernel [37], Katz measure [38-41], the number of paths (PathCount), and the transition matrix of random walk (RW) [42].

According to previous studies, the methods which have some parameters controlling local/global proximity and model the differences among node types, seem superior to the simple measures describe above. For example, Random Walk with Restart on HIN (RWRH) is widely used for biological HIN [43-49]. RWRH uses the multiple domain data, constructs HIN, defines the transition probability within-domain and cross-domain using experimental data or external databases, performs an iterative random walk calculation, and finally outputs the similarity among nodes. Bi-directional RWR [50,51], which is an extension of the RWRH, is also used.

Furthermore, some meta-path based methods have also been applied to biological HIN [42,52-57]. Such approaches explicitly define the different semantic paths. For example, in Fig. 2, multiple paths between Gene Expression and Phenotype, with the limitation of a path length of 2, can be considered, such as Gene Expression–Chemical–Phenotype path, Gene Expression–Microbe–Phenotype path, and so on. Once the similarities of all paths are calculated, any analytical methods are applicable [42,57]. Though PathSim [58] is perhaps the best-known similarity used in meta-path based methods, HeteSim [59] is practically used for biological HIN because HeteSim can be applied to symmetric paths. The methodologies are still developing, such as the appropriate path length [60], informative meta-path selection [61] and weighted summation of each path [62].

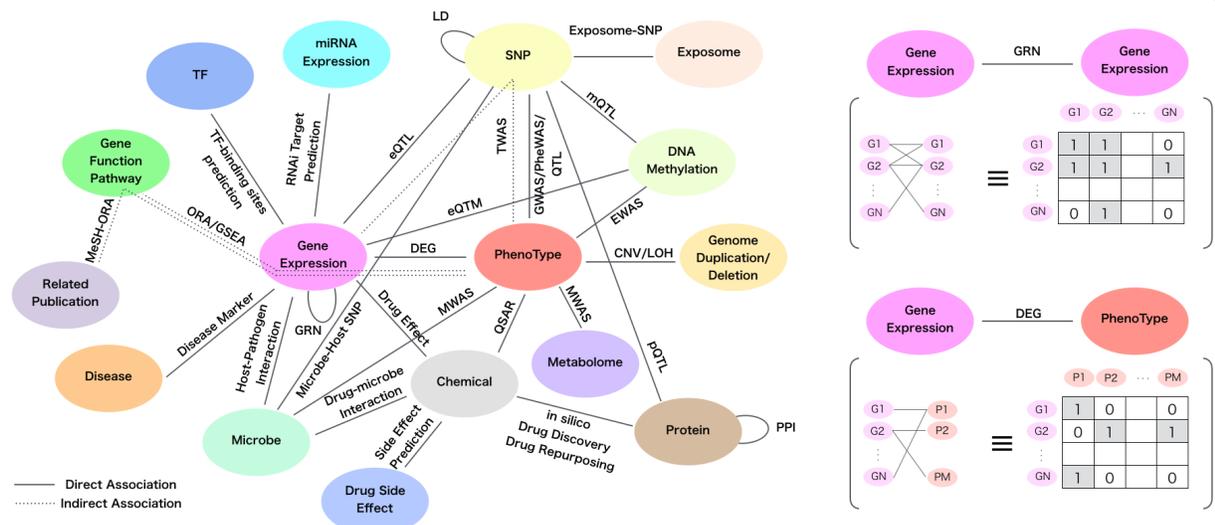

**Figure 2: ER-Diagram of previous omics studies**

Some HIN, such as biomedical RDF data, has several tens of domains that are highly complicated. In such cases, possible combinations of transition matrices (RWRH) or different semantic paths (HeteSim) are exponentially increased and data preparation tasks become very time-consuming and laborious. For simplification of such tasks, some homogeneous information network-based methods, which do not recognize the differences among node types, are directly applied to HIN, such as network embedding methods (https://github.com/chihming/awesome-network-embedding) [63-65].

## 4  WHY IS HIN USEFUL FOR BIOLOGICAL DATA?

The effectiveness of HIN-based methods is featured in some biological research fields. There may be two reasons for that.

Firstly, HIN-based methods enhance the S/N ratio of previous biological data analysis. For example, here, we consider the prediction problem of gene-disease relationships (Fig. 3, upper row).

Although such relationships are measured experimentally by DEG [4] or GWAS [5] study, such data are very noisy and contains many false positives. HIN-based methods can be used not only to measure data but also incorporating previous knowledge about genes/diseases and can seamlessly integrate them. Such data integration will elevate the true link to a higher rank ("prioritization") and can also detect hidden candidate links ("missing-link").

Secondly, HIN-based methods can connect different biological domain data. Again, consider the prediction problem of drug-protein relationships and suppose that there are two data matrices describing drug side-effect relationships and side-effect protein relationships (Fig. 3, lower row). HIN-based methods indirectly connect drug-protein relationship via internal side effect entity, even if the drug-protein relationships are not measured simultaneously.

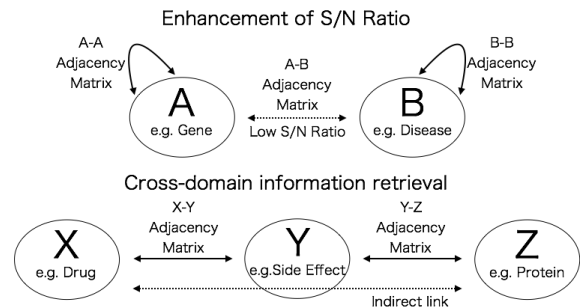

**Figure 3: Effectiveness of HIN-based methods against biological data**

## 5  GUILT-BY-ASSOCIATION

In this section, we explain the principle of "*Guilt-by-Association*" (GBA) [66]. When we applied HIN-based methods to biological data, we will implicitly accept this principle. GBA indicates that biomolecules that are associated or interacting are more likely to share functions. For example, in the case of gene-disease prediction problems, GBA assumes that an unknown gene and a well-known gene are similar and a well-known gene is related to a disease, the unknown gene is also related to the disease (Fig. 4).

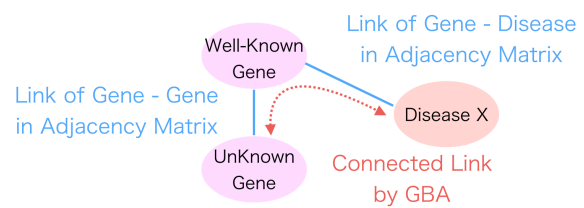

**Figure 4: Concept of Guilt-By-Association**

Integrated pieces of prior knowledge will become a by-pass of information and connect indirect links on HIN that have still not been found. Although this is heuristics rather than rigorous rules, such as a mathematical theorem, many studies suggest the effectiveness as describe below.

## 6    CASE STUDIES OF HIN APPLICATIONS TO BIOLOGICAL DATA

In this section, we describe some previous applications of HIN-based methodology to biological data.

### 6.1    Drug repurposing

The process of finding new drugs is very time-consuming and costs a large amount of money. Drug repurposing (or repositioning) [20] is the application of known drugs to new diseases and is considered a way to resolve this problem; such drugs have already passed a number of toxicity and other tests, omitting significant costs and time to bring a drug to market. Though searching across space for possible combinations of known drugs and diseases is extremely challenging, "*in silico*" (computational) approaches have been used to list the candidate drug-disease pairs.

To find the missing link, an HIN schema, which is the first example in Fig. 3, has often been constructed. Three types of adjacency matrices have been prepared: 1. a drug–drug adjacency matrix, which is constructed based on the similarity among the chemical structures of drugs, the similarity of the related side effects and the semantic similarity of disease ontology; 2. a drug-target adjacency matrix, which is constructed using a prior knowledge database; and 3. a target-target adjacency matrix, which is constructed, using a prior knowledge database of protein-protein interactions (PPI), with sequence similarity as amino acids and semantic similarity of disease-related ontology. Finally, Katz [38] and RWRH [43,44] have been widely applied to find the missing link on the HIN.

More complicated HIN schemas have also been tried. Gang Fu et. al. [52] constructed HIN using 12 types of biological data, calculated 51 paths, implemented a simple random walk measure on each path and evaluated which path is effective in predicting previous drug–target relationships. Congcong Li et. al. [53], used HeteSim [59] against HIN and compared it with previous method based on homogeneous information networks. Other methods for drug–target prediction are reviewed in [67].

### 6.2    Identification of Gene-Phenotype Relationships

Understanding gene function is fundamental to biology, but exhaustive investigation of the function is still difficult. Some systematic approaches have been introduced, such as DEG [4] or GWAS [5] study (Fig. 2). In such approaches, thousands to millions of biomolecules have been investigated simultaneously, with a slight association between a change in quality or quantity of biomolecules and phenotypes. Such experimental designs are intrinsically influenced by experimental noise, resulting in the analysis containing many false positive. To avoid the noise and prioritize the true positives, many HIN-based methods have been developed.

As in 6.1, an HIN schema, which is the first example of Fig. 3, is often constructed. Three type of adjacency matrices are prepared; 1. a gene-gene adjacency matrix, which is constructed using prior knowledge, a PPI database, semantic similarity of gene ontology (GO), similarity of gene expression measured by omics experiments, sequence similarity of amino acids and similarity of related pathways; 2. a gene-disease adjacency matrix, which is constructed using previous knowledge; and 3. a disease-disease adjacency matrix, which is constructed using a semantic similarity of disease related ontology. Finally, Katz [39], RWRH [45,46] and Bi-directional RWR [50] are widely applied.

As more complicated HIN schemas, Xiangxiang Zeng [54] integrated the gene-phenotype relationship of multiple species as HIN, generated a meta-path between gene-human phenotypes and applied HeteSim. Other methods for gene-disease prediction are reviewed in [37].

### 6.3    Functional annotation of non-coding RNA

Non-coding RNA is a generic name for RNAs that are not translated to proteins [15]. Based on its sequence length, cellular location and physical properties, many types of non-coding RNA are categorized. Although such RNAs have been regarded as "junk", recent international genomic projects, such as ENCODE [68] and Epigenome RoadMap [69], have suggested that many non-coding RNAs are functional and the functional annotation is becoming an important problem. In particular, HIN-based methods are applied to miRNA and lncRNA as shown below.

miRNA (micro RNA) is one class of small non-coding RNA that is less than 22 nucleotides. The miRNA-protein complex, which consists of miRNA and its partner proteins, interacts with target mRNA and interferes with the translation (RNA interference or RNAi, Fig. 1). Because only a few target gene sets of human miRNAs have an intersection, functional annotation of miRNA is still a difficult task.

lncRNA (long noncoding RNA) is a class of non-coding RNA that is longer than 200 nucleotides. Some large-scale genomic projects, such as GENCODE [70], detect the expression of some lncRNAs in some tissues and imply a relationship with the tissue-specific functions. However, except for some well-known lncRNA, such as XIST and HOTAIR, most of the lncRNA functions remain unknown.

To leverage such weak non-coding RNA – phenotype (especially disease) links, HIN-based methods, such as Katz [40], RWRH [47,48], HeteSim [55,56] and the meta-path-based logistic regression model [57], have been applied.

### 6.4    Human Microbe – Disease Association

rRNA (ribosomal RNA) is part of a ribosome, which is a protein-RNA complex that is widely used for investigation of the phylogenic relationship among organisms. In particular, 16S rRNA, which is a prokaryote-specific molecule, is used for bacteria research [71]. The relationships between the existence of specific bacterial 16S rRNA and the host human disease have been investigated by the Human Microbe-Disease Association Database (HMDAD) [72], and HIN-based methods have been applied to the data [41,49,51].

## 7 ADVANCED TOPICS

In this section, we introduce some advanced topics for future biological data analysis with HIN. Owing to the outstanding property of HIN-based methods described above, many biological problems can be solved.

### 7.1 Biological Domain Specific Topics

Generally, biological data are noisy, and sometimes the value is systematically lost. Therefore, missing-value estimation methods (imputation) have been developed in each biological domain. HIN-based methods may work for imputation problems; imputation is theoretically similar to collaborative filtering [73], which is known as the methodology used in item-recommendation systems, such as Amazon and Netflix and HIN-based methods, are still applied to collaborative filtering problems [74].
 When using omics data, a large number of biomolecules are simultaneously measured. Generally, the number of biomolecules related to phenotypes is considered small, and hypothetical tests are applied and significant molecules are listed. Some hypothetical tests on HIN have been prepared such as the hypergeometric test [48], Wilcoxon rank-sum/Kruskal-Wallis test [75] and Permutation test [76].
   Furthermore, it is natural to think that the relationships among different biological data are many-to-many, rather than 1-to-1. Such relationships can be detected by a latent variable model such as matrix factorization methods [77]. Actually, the restricted Boltzmann machine, which is one of the latent variable models, successfully detects multi-lncRNA-multi-diseases relationships [78].

   In the biological research community, many end-users of these analytical methods are not programmers but experimental researchers. Therefore, these methods should be implemented in a format that users can easily try on their own data. In addition to the release from the GitHub repository, implementation in a Bioconductor (R) package is desirable; many omics analytical methods are implemented and an analytical pipeline is implemented in combination with multiple R packages (e.g., https://bioconductor.org/help/workflows/). Some HIN-based methods are still implemented as R packages [79,80].

### 7.2 More Specific, Noisy and Sparse Data

Although previous applications of HIN-based methods to biological data have focused on an entity such as drugs and genes (see Section 6), these methods are also effective against more specific data because the link between such data and previous knowledge is considered to be weak due to the small number of studies. For example, studies have used data from some non-coding RNA such as circRNA and eRNA, long range interaction of SNP-gene expression (trans-eQTL) [81], phosphorylation [82], non-model organisms and cross-species analysis [83], orphan and rare disease [67], traditional medicine [84], and so on.
   Some biological data, such as epigenomic data (Fig. 1) [2,25,27,28] and brain MRI images [85], are highly noisy and sparse. Because such data have low S/N ratios, the HIN-based method may work for such data.

### 7.3 Higher Resolution Data

The technological innovation of a high-resolution measurement of a biological system is rapidly progressing. For example, previous experiments by disease researchers were designed to make comparisons between patients' groups and control groups, but recent personalized medicine [86] focuses on each patients' heterogeneity. In the same way, a single-type omics will become multi-omics resolution [3,10], gene expression measurement of both alleles will become allele specific expression [87], a single gene-level analysis will become isoform-resolution [88], and a microbe analysis based on 16S rRNA will become a Dual RNA-Seq [11], which can focus on the gene expression of pathogen and host cells simultaneously.

### 7.4 Single-Cell Omics and Human Cell Atlas

Typical omics experiments have been applied to samples consisting of thousands or millions of cells. The recent rapid development of DNA sequence technologies and cell capture technologies enable us to perform experiments at a single-cell level (single-cell omics [89]) such as single-cell genome, transcriptome, epigenome, proteome, metabolome, single-cell multi-omics [90] and single-cell dual RNA-Seq [91]. These ultimately high-resolution experiments identify novel cell types consisting of tissues [92], changes in gene expression along with cell development [93] and spatial expression patterns [94]. In spite of some success, single-cell data are very noisy and sparse, due to the small number of biomolecules within single-cell. The HIN-based method will enhance the S/N ratio of links between each molecules and cell types and will integrate different types of omics data.

   The human cell atlas (https://www.humancellatlas.org), which is funded by the Chan Zuckerberg Initiative (https://chanzuckerberg.com), was started in 2016 [95]. The atlas will perform single-cell RNA-Seq against 30 million to 10 million cells, categorize cell types from the human body, and investigate some mechanisms of disease at the single-cell level. The atlas data will become a reference for interpretation of further generating other types of single-cell omics data. HIN-based methods will seamlessly integrate these different domain data, such as single-cell Epigenome-single-cell RNA-Seq-Gene Function path, and will support interpretation of the causal relationships among them, within common cellular space.

### 7.5 More Complicated Data Structure

Many HIN-based methods are based on an adjacency matrix, which means that each element has 1 (linked) or 0 (not linked).

However, some biological relationships have a plus direction (e.g., up-regulated genes) and minus direction (e.g., down-regulated genes). Such signed graph relationships are defined as three values (-1: negative, 0: not linked, and 1: plus). Furthermore, some data are generated from spatiotemporally different sample points. Because such data structures are defined as multiple matrices or tensor, tensor-based methods are needed [36].

More sophisticated algorithms, such as graph convolutional neural network [96], which is the graph extension of the deep learning method or metapath2vec [97], which is a combination of meta-paths and network embedding methods, may work for a graph with a complicated network schema.

### 7.6 More General Framework and Community-Driven Knowledge Sharing

Some HIN-based software, such as SLAP [98], Hetionet [60], Medusa [77], and DeepWalk implemented at Biohackathon 2016 [64], integrate a large variety of domain data. In principle, any connection between start-node to end-node can be linked in any combination. Therefore, more general and unified frameworks may be able to be implemented, wherein related data are assigned to query data without a consideration of different data domains, at single search interface (Fig. 5).

There are still some problems to address. Firstly, the best practices of data analysis workflow in each data domain are not always determined. Therefore, the data generated by different laboratories may not be directly comparable, if the software used is different. Therefore, to standardize the analytical results of data from a primary database, secondary databases for re-analyzing the data and providing them are essential.

Secondly, the vocabulary of biology is not always unique. For example, POU5F1, which is a famous gene for establishing the iPS cell, has some synonyms, such as MGC22487, Oct3, and Oct4 (https://www.genenames.org/cgi-bin/gene_symbol_report?hgnc_id=HGNC:9221). Therefore, to integrate the different laboratory and different domains, the designation of common ontology is needed.

As we described above, there are wide variety of biomolecules and data domains, the biological databases assigned different biomolecular types to different types of IDs such as Gene ID and Protein ID. For such a reason, ID conversion across different biological domains is laborious task. In view of this situation, some biological specific programming framework such as BioPerl, BioPython, BioRuby and BioJulia provided ID conversion functions via Web-API. To the contrary, Bioconductor provided the corresponding ID tables as R packages. Recently, semantic web technologies have been applied to biological databases. Some databases are already reconstructed in RDF format and the data can be seamlessly searched without consideration of difference of data domains [64]. However, the query language of Semantic Web (SPARQL) can only retrieve corresponding data and SPARQL cannot perform prioritization or missing-link prediction, which is described in this work. Therefore, combined use of SPARQL and HIN-based methods will be considered, along with the differences in user purpose.

## 8 CONCLUSIONS

In this work, we introduced some applications of HIN-based methods to biological data and described some future research directions. Many biological data analyses are predictions of associations between two entities, and the outstanding property of HIN-based methods will be preferable for noisy, sparse, and complicated biological data and leverage some unresolved biological problems.

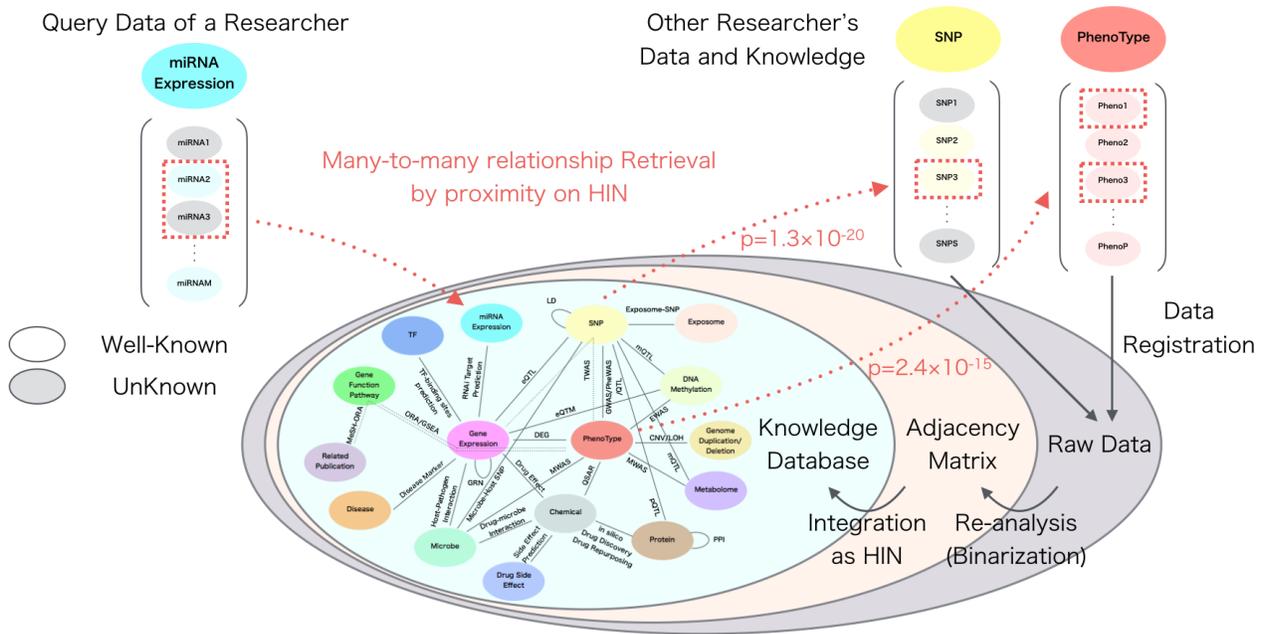

**Figure 5: Future Biological Data Analysis**

## ACKNOWLEDGMENTS

This work was supported by JSPS KAKENHI, Grant Number 16K16152. Some images used in the Figure 1 were provided by DBCLS Togo Picture Gallery (© 2016 DBCLS TogoTV / CC-BY-4.0).

## Appendix: Abbreviation List

circRNA: Circular Ribonucleic Acid
CNV: Copy Number Variation
DEG: Differentially Expressed Gene
DNA: Deoxyribonucleic Acid
eQTL: Expression Quantitative Trait Loci
eQTM: Expression Quantitative Trait Methylation
eRNA: Enhancer Ribonucleic Acid
EWAS: Epigenome-wide Association Study
GBA: Guilt-By-Association
GRN: Gene Regulatory Network
GSEA: Gene Set Enrichment Analysis
GWAS: Genome-wide Association Study
LD: Linkage Disequilibrium
lncRNA: Long non-coding Ribonucleic Acid
LOH: Loss of Heterozygosity
MeSH-ORA: Medical Subject Headings - Over-representation Analysis
miRNA: Micro Ribonucleic Acid
mQTL: Metabolomic Quantitative Trait Loci or Methyl Quantitative Trait Loci
MWAS: Metabolome-wide Association Study or Microbe-wide Association Study
ORA: Over-representation Analysis
PheWAS: Phenome-wide Association Study
PPI: Protein-Protein Interaction
pQTL: Protein Quantitative Trait Loci
QSAR: Quantitative Structure Activity
QTL: Quantitative Trait Loci
RDF: Resource Description Framework
RNA: Ribonucleic Acid
rRNA: Ribosomal Ribonucleic Acid
SNP: Single Nucleotide Polymorphism
TF: Transcriptional Factor
TWAS: Transcriptome-wide Association Study